\documentclass{PoS}
\usepackage{epsfig}
\def\gsi{\raise0.3ex\hbox{$>$\kern-0.75em\raise-1.1ex\hbox{$\sim$}}}

\newcommand{\beq}{\begin{equation}}
\newcommand{\eeq}{\end{equation}}
\newcommand{\beqn}{\begin{eqnarray}}
\newcommand{\eeqn}{\end{eqnarray}}
\newcommand{\bi}{\begin{itemize}}
\newcommand{\ei}{\end{itemize}}
\newcommand{\benu}{\begin{enumerate}}
\newcommand{\eenu}{\end{enumerate}}
\newcommand{\tr}{{\rm Tr\,}}
\newcommand{\psibar}{\bar\psi}

\title{The two flavour Schwinger model: \\
scaling of the scalar condensate}

\ShortTitle{The two flavour Schwinger model}

\author{\speaker{Kei-ichi Nagai}\\
ZAM, Forschungszentrum J\"ulich, 52425 J\"ulich, Germany\\
FB-C/Physik, Bergische Universit\"at Wuppertal, 42097 Wuppertal, Germany\\
E-mail: \email{knagai@theorie.physik.uni-wuppertal.de}
}

\author{Nils Christian\\
Theoretical Biophysics, Humboldt University at Berlin,
Invalidenstr. 42, 10115 Berlin, Germany\\
E-mail: \email{nils.christian@physik.fu-berlin.de}
}

\author{Karl Jansen\\
DESY, Zeuthen, Platanenallee 6, 15738 Zeuthen, Germany\\
E-mail: \email{Karl.Jansen@desy.de}
}

\author{Beatrix Pollakowski\\
Physikalisch-Technische Bundesanstalt (PTB),
Abbestr. 2-12, 10587 Berlin, Germany \\
E-mail: \email{beatrix@physik.hu-berlin.de}
}

\abstract{
We investigate the continuum limit scaling of the scalar condensate
in the $N_f=2$ Schwinger model on the lattice.
We employ maximally twisted mass Wilson fermions and overlap fermions.
We compute the scalar condensate
by taking the trace of the propagator (direct method)
and by utilizing the integrated Ward-Takahashi identity.
While the scalar condensate comes out consistent
using these two methods for a given kind of lattice fermions,
we find --quite surprisingly-- large discrepancies for the scalar condensate
between twisted mass and overlap fermions.
These discrepancies are only resolved
when using the point split current for twisted mass fermions.
}

\FullConference{The XXV International Symposium on Lattice Field Theory\\
                 July 30-4 August 2007\\
                 Regensburg, Germany}

\begin{document}

\section{Introduction}
\label{sec:intro}

The Schwinger model~\cite{Schwinger:1962tp}
is a good test ground for 4-dimensional QCD
due to the properties of
asymptotic freedom and
the existence of non-perturbatively generated bound states.
In refs.~\cite{Christian:2005yp}
we have studied on a lattice the scaling properties of meson
masses for a number of different fermion discretizations 
using Wilson, hypercube, Wilson twisted mass (TM) and 
overlap (OV) fermions.
In these investigations, we found that
the pseudo scalar mass scales with an $O(a^2)$ behaviour
for all fermion discretizations.
The same $O(a^2)$ scaling behaviour was also observed in
\cite{DellaMorte:2005vc} in a finite
volume scaling analysis in the Schwinger model.
These findings can be attributed to the super-renormalizability of the model.

The aim of the present work is to perform another scaling test 
of the scalar condensate as an additional, non-trivial quantity.
In the case of $N_f=1$,
ref.~\cite{Durr:2004ta} shows that the continuum and the chiral limit
come out to be consistent with the predictions 
in the continuum theories~\cite{Hetrick:1995wq,Smilga:1996pi}.
Here we investigate the scaling toward the continuum limit 
in the $N_f=2$ Schwinger model
using maximally twisted mass fermions
and overlap fermions.
Emphasis will be put on the comparison of
different methods and definitions
of the scalar condensate used for its computation.
The methods we apply and which will be detailed below are
the direct calculation through the trace of the (inverse) Dirac operator
employed and the integrated axial Ward--Takahashi identity.
As a quite surprising outcome of this investigation we find that a naive
approach of computing the scalar condensate by using the
local definition, $\langle \bar{\psi}(x)\psi(x)\rangle$,
does not lead to a consistent continuum limit.

Another aspect of our work which was presented in the poster
has been the
investigation of the question of how many eigenmodes 
of the used lattice Dirac operator are needed to
approximate the pseudo scalar correlator to a certain precision.
In particular, we have studied how this number of
eigenmodes scales towards the continuum limit
in the case of overlap fermions.
We can, for lack of space, not discuss this issue in this proceedings 
write-up and refer to ref.~\cite{Nagai:2007} for a detailed discussion.

\section{Lattice actions and calculation methods}
\label{sec:method}

We have employed  maximally twisted mass Wilson fermions   
and overlap fermions as chirally improved and chirally invariant formulations
of lattice fermions, respectively, in this work.
Since both of these kind of lattice fermions 
are $O(a)$ improved,
we expect only $O(a^2)$ lattice artefacts.

The Neuberger operator~\cite{Neuberger:1997fp} as a realization of
overlap fermions is given as
\begin{equation}
D_{\rm ov} = \left( 1 - \frac{m_q a}{2} \right) D_0 + m_q \quad , \quad
D_0 = \frac{1}{a} \left[ 1 + \frac{D_{\rm kernel}}{\sqrt{D_{\rm kernel}^{\dagger} D_{\rm kernel}}} \right] \ .
\label{eq:overlap}
\end{equation}
In the following,
we will use as kernel the hypercube operator~\cite{Bietenholz:1999km},
i.e. $D_{\rm kernel} =D_{\rm hyp}$ with parameters obtained from optimizing scaling.
This lattice fermion has an exact (lattice) chiral symmetry 
due to the Ginsparg--Wilson relation~\cite{Ginsparg:1981bj}.


For Wilson twisted mass fermions, the
lattice Dirac operator is given by
\begin{equation}
D_\mathrm{tm}(x,y)= (m_0 + 2) \delta_{x,y} -
      \frac{1}{2}\sum_{\mu=1}^2[(1-\sigma_\mu) U_\mu(x)
      \delta_{x,y-\hat\mu} +
      (1+\sigma_\mu)U^\dag_\mu(y) \delta_{x,y+\hat\mu} ]+  i \mu_\mathrm{tm} \sigma_3 \tau_3 \delta_{x,y}\; .
\label{eq:tmaction}
\end{equation}
We use the standard Pauli-matrices $\sigma_\mu$ $(\mu=1,2,3)$ with
$\sigma_3=\mathrm{diag}(1,-1)$ and denote with $\hat\mu$ the unit vector
shift in direction $\mu$.
The parameters $m_0$ and $\mu_\mathrm{tm}$ denote the untwisted and twisted bare
fermion masses, respectively.
When $m_0$ is tuned
to a critical value, $m_0=m_{0,\mathrm{crit}}$, by tuning the PCAC 
quark mass to zero,
we reach maximally twisted mass fermions~\cite{Frezzotti:2003ni}. 
In this case, the theory is automatic
$O(a)$-improved and physical (parity even) quantities
scale with an $O(a^2)$ behaviour toward the continuum limit.

There are two methods on the lattice,
the direct and the integrated Ward--Takahashi identity methods, 
to calculate the scalar condensate which we will use
for overlap and maximally twisted mass fermions.

The direct method is obtained
from the trace of the fermion propagator
with gauge backgrounds.
The direct method is defined as follows
for the overlap and twisted mass fermions respectively;  
\begin{equation}
\Sigma_{\rm direct}^\mathrm{ov}  = \frac{1}{V} \sum_x \tr \left[ (1 - \frac{a D_0}{2})D_{\rm ov}^{-1} \right]_{(x,x)}  \quad , \quad
\Sigma_\mathrm{direct}^\mathrm{tm} =\frac{1}{V} \sum_x \tr \left[ i \sigma_3 \tau_3 D_{\rm tm}^{-1} \right]_{(x,x)} \; .
\label{eq:direct}
\end{equation}
 
For Wilson fermion,
there is a term proportional to $\frac{1}{a}$ in 2-dimensions 
due to the explicit breaking 
of chiral symmetry~\cite{Bochicchio:1985xa}.
However overlap fermions and maximally twisted mass fermions
do not have such a problem
and thus no $\frac{1}{a}$ term in 2-dimensions~\cite{Christian:2005yp} appears. 
The same is true for the $\frac{1}{a^3}$ term in 4-dimensions.

Next we introduce the integrated Ward--Takahashi identity method~\cite{Bochicchio:1985xa} which 
reads
\beq
\Sigma_{\rm iWT}^\mathrm{ov} = 2 m_q \sum_x \langle P^{+}(x) P^{-}(0) \rangle \quad , \quad
\Sigma_{\rm iWT}^\mathrm{tm} = 2 \mu_\mathrm{tm} \sum_x \langle P^+(x) P^-(0) \rangle  \;.
\label{eq:iwt}
\eeq
$\Sigma_{\rm iWT}^\mathrm{ov}$ is obtained from the PCAC relation with the operator
$P^{\pm}(x) = \psibar(x) \sigma_3 \tau_{\pm} [(1 - \frac{a D_0}{2}) \psi](x)$ 
and $\Sigma_{\rm iWT}^\mathrm{tm}$ from PCVC relation for TM fermions;
$\partial^{*}_\nu \langle V^{+}_\nu(x) P^{-}(0) \rangle=
2 \mu_\mathrm{tm}  \langle P^{+}(x) P^{-}(0) \rangle-\delta_{x,0} \langle S_0(x) \rangle $\\
where
$P^{\pm}(x)=\psibar(x) \sigma_3 \tau_{\pm} \psi(x)$, $S_0(x)=i \psibar(x) \sigma_3 \tau_3 \psi(x)$.

\section{Calculating the scalar condensate}
\label{sec:condensate}

We have carried out numerical simulations in the following setting.
The lattice size is $20 \times 20$ and 
the statistics is over 1000 thermalized and statistically 
independent configurations.
The quark mass is fixed as $z=\left( m_q \sqrt{\beta} \right)^{2/3}=0.4$.
The error estimate is done by the method of ref.~\cite{Wolff:2003sm}.
The gauge action ($S_G$) is the Wilson plaquette action
with the dimensionless coupling constant $\beta=\frac{1}{e^2 a^2}$.

To obtain results for dynamical fermions, 
for OV fermions, we simulated only the gauge action and 
used the re-weighting method and the spectral representation
from eigenvalues and eigenmodes to compute physical observables:
$\langle {\cal O} \rangle_{\rm unquench}
= \frac{\langle \det^{N_f}(D_f) \cdot {\cal O} \rangle_{S_G}}{\langle \det^{N_f}(D_f) \rangle_{S_G}} $.
For TM fermions,
we also generated configurations for the full action
using the HMC algorithm.

\begin{figure}[h!]
\begin{center}
\resizebox{12cm}{!}{\rotatebox{0}{\includegraphics{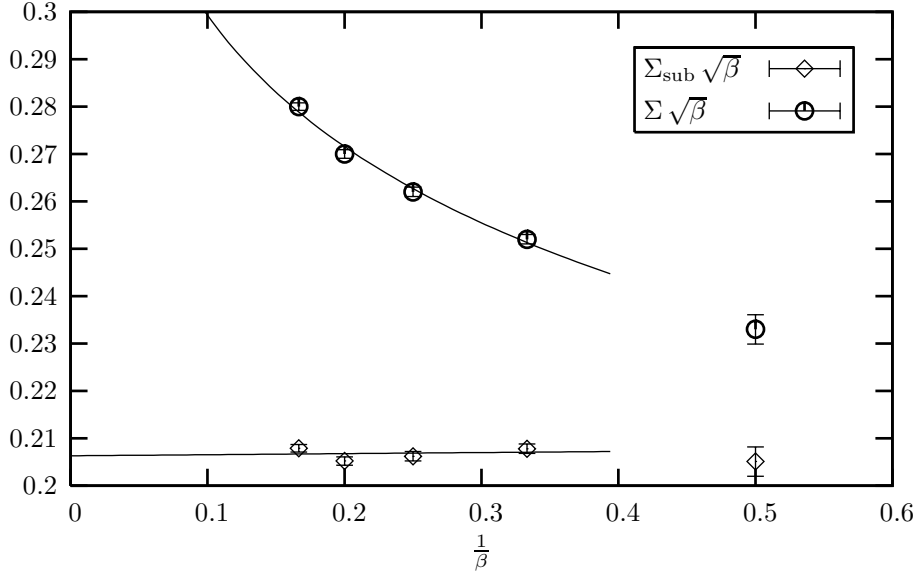}}}
\caption{
$\sqrt{\beta}\Sigma$ and $\sqrt{\beta}{\Sigma}_\mathrm{sub}$ in the case of OV fermions
as a function of $1/\beta$ for a fixed value of $z=0.4$.}
\label{fig:sigmasub}
\end{center}
\end{figure}

We want to remark first that all results obtained by using 
either 
$\Sigma_{\rm direct}^\mathrm{ov}$ 
($\Sigma_{\rm direct}^\mathrm{tm}$) 
or $\Sigma_{\rm iWT}^\mathrm{ov}$,
($\Sigma_{\rm iWT}^\mathrm{tm}$)
came out to be completely consistent when considered for each kind of 
lattice fermion separately.
Therefore,
we discuss in the following only one of these cases.
From a dimensional analysis, we expect a 
logarithmic term in $\beta$, \cite{Durr:2004ta,Christian:2005yp}
\beq
\sqrt{\beta}\Sigma = A + B/\beta + C\log(\beta)\; .
\label{eq:logscaling}
\eeq
Because of 
universality 
and super-renormalisability,
the coefficient multiplying the logarithmic term 
is universal 
and can be evaluated
as $C=\frac{m_q\sqrt{\beta}}{2 \pi}$.
This allows us to define a subtracted scalar condensate,
\beq
\sqrt{\beta}{\Sigma}_\mathrm{sub} = \sqrt{\beta}\Sigma - C \log(\beta) \; .
\label{eq:logsub}
\eeq
Fig.~\ref{fig:sigmasub} shows 
$\Sigma\sqrt{\beta}$ and 
${\Sigma}_\mathrm{sub} \sqrt{\beta}$ 
in the case of OV fermions. 
When the logarithmic term
is subtracted, we observe a perfectly linear behaviour 
of ${\Sigma}_\mathrm{sub} \sqrt{\beta}$ as a function 
of $1/\beta$.   

In order to tune to maximal twist we have determined 
the critical Wilson mass in the 
pure Wilson theory, i.e. setting
$\mu_\mathrm{tm}=0$, by tuning the (untwisted) PCAC quark mass to zero.
The values of the $m_{0,\mathrm{crit}}$ can be found in ref.~\cite{Christian:2005yp}.

\begin{figure}[h!]
\centering
\centerline{
\resizebox{12cm}{!}{\rotatebox{0}{\includegraphics{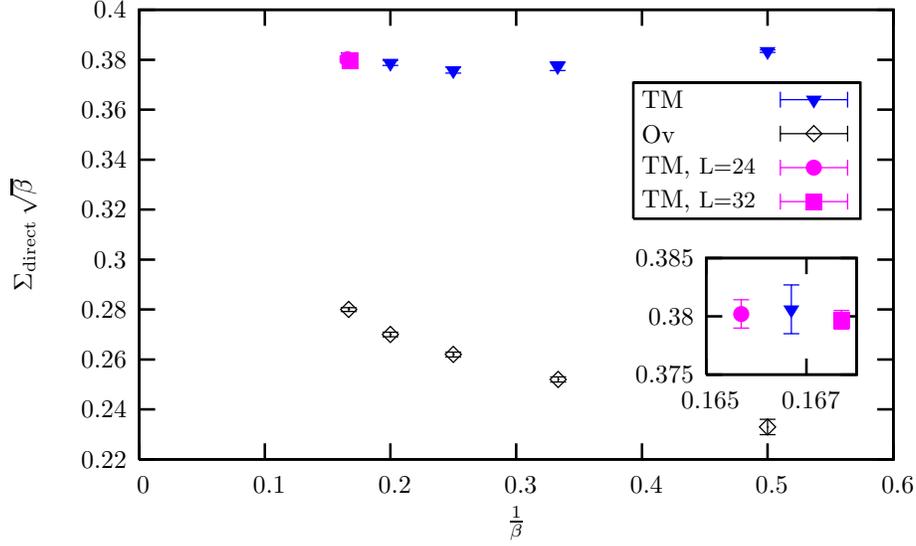}}}
}
\caption{Comparison of the scalar condensate using OV and TM fermions 
as a function of $1/\beta$.
In the inlet data on larger lattices are shown, demonstrating
the smallness of finite size effects.}
\label{fig:cond}
\end{figure}

In fig.~\ref{fig:cond},
we compare the (direct) TM fermion scalar condensate
with the corresponding (direct) OV fermion data.
The important result is
that $A_{\rm tm} \neq A_{\rm ov}$, see eq.~(\ref{eq:logscaling}). 
Thus the continuum extrapolated values of the scalar 
condensate from both discretizations differ when the 
fitting function of eq.~(\ref{eq:logscaling}) is used.

In order to shed some light on the discrepancy between the 
results for TM and OV fermions discussed above, 
we suggest to use an improved current for TM fermions, namely
the 1-point splitting current, given as
\beq
S_\mathrm{tm}^\mathrm{imp}(x) = \frac{1}{d} \sum_{\mu=1}^d \frac{1}{2}
\left[ \psibar(x) \gamma_5 \tau_3 U_\mu(x) \psi(x+\hat{\mu})
+ \psibar(x) \gamma_5 \tau_3 U^\dagger_\mu(x-\hat\mu) \psi(x-{\hat\mu}) \right] \;  
\label{eq:d-imp}
\eeq
for the direct method. For the 
integrated Ward--Takahashi identity we find 
\beq
\partial^{*}_\nu \langle V^{+}_\nu(x) \hat{P}^{-}(0) \rangle
= 2 \mu_\mathrm{tm} \langle P^{+}(x) \hat{P}^{-}(0) \rangle  + \delta_{PCVC} \langle \hat{P}^{-}(0) \rangle \;,
\label{eq:wt-imp}
\eeq
where $\delta_{PCVC}$ is the chiral rotation in the twisted mass formulation and
\beq
\hat{P}^{\pm}(y) = \frac{1}{d} \sum_{\mu=1}^d \frac{1}{2}
\left[ \psibar(y) \gamma_5 \tau_{\pm} U_\mu(y) \psi(y+\hat{\mu})
+ \psibar(y) \gamma_5 \tau_{\pm} U^\dagger_\mu(y-\hat{\mu}) \psi(y-\hat{\mu}) \right] \;. 
\label{eq:p-imp}
\eeq

\begin{figure}[h!]
\centering
\centerline{
\resizebox{12cm}{!}{\rotatebox{0}{\includegraphics{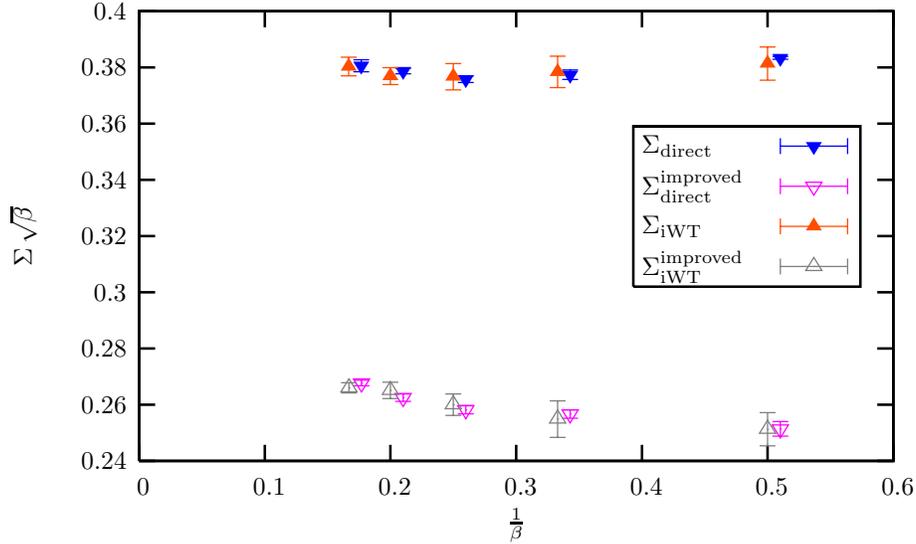}}}
}
\caption{
The scalar condensate for TM fermions
as a function of $1/\beta$,
using both, un-improved and improved currents.
}
\label{fig:cond-tm}
\end{figure}

\begin{figure}[h!]
\centering
\centerline{
\resizebox{12cm}{!}{\rotatebox{0}{\includegraphics{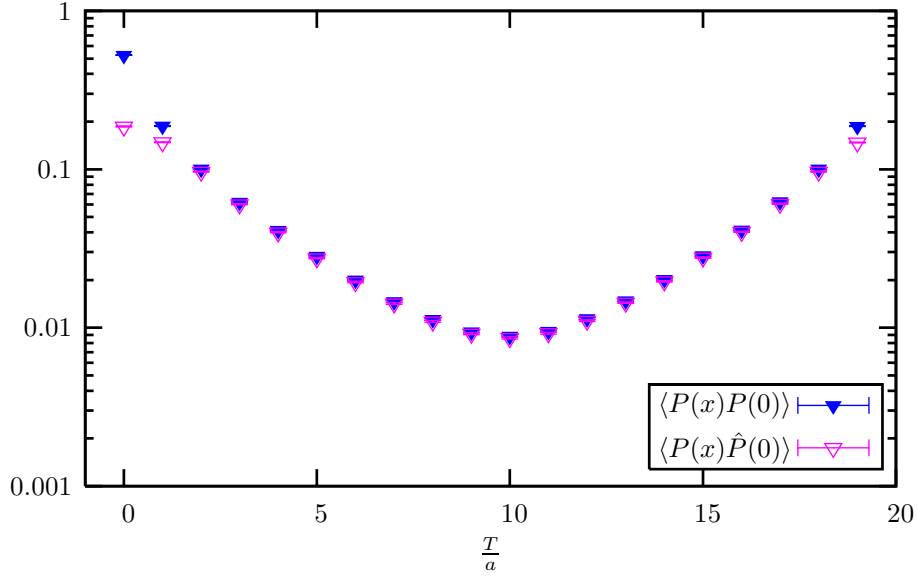}}}
}
\caption{The correlator $\langle P(t) \hat{P}(0) \rangle$ and
$\langle P(t) P(0) \rangle$ for twisted mass fermions. }
\label{fig:prop}
\end{figure}

Fig.~\ref{fig:cond-tm} shows the comparison
using the improved currents in eqs.~(\ref{eq:d-imp},~\ref{eq:p-imp})
with the un-improved currents in eqs.~(\ref{eq:direct},~\ref{eq:iwt}).
We find that employing the improved current, the 
values of the scalar condensate from TM and OV fermions 
approach each other. 
The effect of the improved current is 
illustrated in fig.~\ref{fig:prop},
which shows the correlator of the pseudo-scalar meson  
using the un-improved and the improved currents for the case of 
TM fermions.
It can be clearly observed that the short distance behaviour 
is significantly altered leading to the different behaviour 
of the scalar condensate. 

\begin{figure}[h!]
\centering
\centerline{
\resizebox{12cm}{!}{\rotatebox{0}{\includegraphics{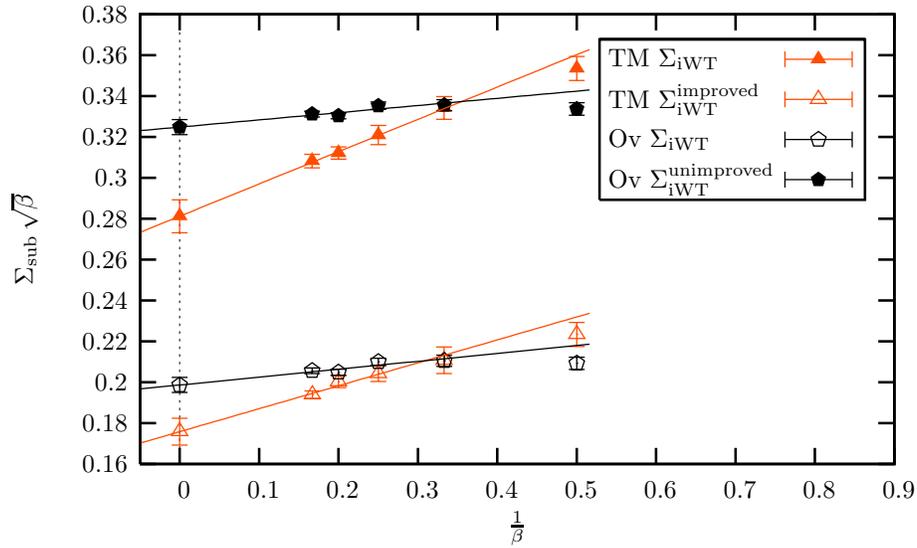}}}
}
\caption{The subtracted condensate for fixed $z=(m_q \sqrt{\beta})^{2/3}=0.4$}
\label{fig:logsub}
\end{figure}

In fig.~\ref{fig:logsub},
$\sqrt{\beta}\Sigma_\mathrm{sub}$ of eq.~(\ref{eq:logsub}) is plotted
as a function of  $1/\beta$.
From this figure, we see that 
$\sqrt{\beta}\Sigma_\mathrm{sub}$ shows the expected 
$O(a^2)$ scaling.
In addition, the results for TM fermions using the improved current
is very close to the one obtained from OV fermions. 
However, the continuum limit values of 
the scalar condensate are still not fully consistent.

\section{Summary}
\label{sec:summary}

We have performed numerical simulations in the 
2-flavour Schwinger model thought as a test ground for lattice QCD.
In our work,
the re-weighting, the HMC algorithm for the calculation of
the fermion determinant 
and the spectral representation for the correlation function
are used.

We carried out a scaling test of the scalar condensate
for OV and TM fermions.
Quite unexpectedly, we found a large discrepancy 
for the values of the scalar condensate between TM 
and OV fermions, even in the continuum limit. 
The discrepancy is particularly significant when an 
un-improved (local) current is used for TM fermions. 
We have suggested to use a 1-point splitting current 
which indeed improves the situation but 
does not seem to remove the discrepancy completely.

As a curious observation, we remark that we have also computed 
the scalar condensate by using an un-improved 
definition for OV fermions defined by 
$\widetilde{P}(x) = \psibar(x) \gamma_5 \psi(x)$,
in which there is no factor $(1 - \frac{a D_0}{2})$. In this case,
the value for the scalar condensate is almost the same value as the 
one obtained from the un-improved current in the case of TM fermions.

The reasons for the quite surprising outcomes of our investigation 
are presently explored. 
Clearly, if our findings are not a specialty of the 2-dimensional 
Schwinger model but generalize to lattice QCD, this would have serious
implications for the determination of the scalar condensate in QCD.

\vspace{2mm}
{\bf Acknowledgments} \  
We thank C.~H\"olbling, G.C.~Rossi, A.~Shindler and
S. Sch\"afer for many useful discussions.
We are grateful to K.M.J.~Adamiak and R.G.~Kuiper
for their initial work on the eigenvalue saturation during their time
as summer students at DESY--Zeuthen.


\vspace*{-2mm}

\begin{thebibliography}{99}

\bibitem{Schwinger:1962tp}
J.~S. Schwinger,
\newblock Phys. Rev. {\bf 128} (1962) 2425.

\bibitem{Christian:2005yp}
N.~Christian, K.~Jansen, K.~Nagai and B.~Pollakowski,
\newblock Nucl. Phys. {\bf B739} (2006) 60,
\newblock PoS {\bf LAT2005} (2006) 239.

\bibitem{DellaMorte:2005vc}
M.~Della~Morte and M.~Luz,
\newblock Phys. Lett. {\bf B632} (2006) 663.

\bibitem{Durr:2004ta}
S.~Durr and C.~Hoelbling,
\newblock Phys. Rev. {\bf D71} (2005) 054501.

\bibitem{Hetrick:1995wq}
J.~E. Hetrick, Y.~Hosotani and S.~Iso,
\newblock Phys. Lett. {\bf B350} (1995) 92.

\bibitem{Smilga:1996pi}
A.~V. Smilga,
\newblock Phys. Rev. {\bf D55} (1997) 443.

\bibitem{Nagai:2007}
N.~Christian, K.~Jansen, K.~Nagai and B.~Pollakowski,
In preparation.

\bibitem{Neuberger:1997fp}
H.~Neuberger,
\newblock Phys. Lett. {\bf B417} (1998) 141.


\bibitem{Bietenholz:1999km}
W.~Bietenholz and I.~Hip,
\newblock Nucl. Phys. {\bf B570} (2000) 423.

\bibitem{Ginsparg:1981bj}
P.~H. Ginsparg and K.~G. Wilson,
\newblock Phys. Rev. {\bf D25} (1982) 2649.

\bibitem{Frezzotti:2003ni}
R.~Frezzotti and G.~C. Rossi,
\newblock JHEP {\bf 08} (2004) 007.

\bibitem{Bochicchio:1985xa}
M.~Bochicchio, L.~Maiani, G.~Martinelli, G.~C. Rossi and M.~Testa,
\newblock Nucl. Phys. {\bf B262} (1985) 331.

\bibitem{Wolff:2003sm}
ALPHA, U.~Wolff,
\newblock Comput. Phys. Commun. {\bf 156} (2004) 143.


\end{thebibliography}
\end{document}